\documentclass[twocolumn,english,preprintnumbers,amsmath,amssymb,pre]{revtex4}

\usepackage{graphicx}
\usepackage{color}

\begin{document}

\title{Isotopic effects in structural properties of graphene}
\author{Carlos P. Herrero}
\author{Rafael Ram\'irez}
\affiliation{Instituto de Ciencia de Materiales de Madrid,
         Consejo Superior de Investigaciones Cient\'ificas (CSIC),
         Campus de Cantoblanco, 28049 Madrid, Spain }
\date{\today}

\begin{abstract}
Isotopic effects are relevant to understand several properties 
of solids, and have been thoroughly analyzed along the years.
These effects may depend on the dimensionality of the considered
solid.  Here we assess their magnitude for structural
properties of graphene, a paradigmatic two-dimensional material.
We use path-integral molecular dynamics simulations, a well-suited 
technique to quantify the influence of nuclear
quantum effects on equilibrium variables,
especially in cases where anharmonic effects are important.
Emphasis is put on interatomic distances and mean-square
displacements, as well as on the in-plane area of the graphene 
layer.  At low temperature, the relative difference in 
C--C distance for $^{13}$C and $^{14}$C, with respect to $^{12}$C, 
is found to be $-2.5$ and $-4.6 \times 10^{-4}$, respectively,
larger than in three-dimensional carbon-based materials
such as diamond.  For the in-plane area, the relative changes 
amount to $-3.9$ and $-6.9 \times 10^{-4}$.
The magnitude of anharmonicity in the lattice vibrations is 
estimated by comparing the internal 
energy and atomic vibrational amplitudes with those derived from
a harmonic approximation.
\end{abstract}

\maketitle

\section{Introduction}

In the last years there has been a surge of interest in carbon-based
materials, especially in those composed of C atoms with $sp^2$
hybridization, such as fullerenes, carbon nanotubes,
and graphene, a two-dimensional (2D) crystal with extraordinary
electronic \cite{ge07,fl11}, elastic \cite{le08}, and 
thermal properties \cite{gh08b,ni09,ba11}.
The structural pattern for pure defect-free graphene
corresponds to a honeycomb lattice, but
deviations from this flat structure can appreciably affect
its atomic-scale and macroscopic properties \cite{me07}.
In fact, thermal fluctuations at finite temperatures give rise to
out-of-plane vibrations of the C atoms, and for $T \to 0$ the 
graphene sheet cannot be strictly planar due to zero-point motion.

Anharmonic effects in condensed matter have been studied since many
years, because they are responsible for important effects
such as thermal expansion, pressure dependence of the
compressibility, and phonon couplings, along with the isotope
dependence of structural properties and melting
temperature \cite{as76,ki96,ra10}.
In this line, the effect of isotopic composition on the electronic 
properties, structural parameters, and lattice dynamics of 
three-dimensional (3D) materials, as well as low-dimensional 
compounds has been studied with great detail \cite{ca93,ca00,he09c,he11}.
More recently, several isotopic effects have been evaluated
in graphene by using various techniques,
e.g., vibrational spectroscopy \cite{be12,br14}
and in particular Raman scattering \cite{ro12,co13,ca15b}.
Moreover, detailed studies about the effect of the atomic mass on 
the thermal conductivity of graphene have been reported by several 
research groups \cite{hu10,ji10,ad12,da17},

Different types of isotopic effects have been analyzed in materials,
mainly those caused by the change of phonon frequencies with the
atomic mass \cite{ca00}. This mass dependence of the frequencies
gives rise to variations in the vibrational amplitudes. Although at
high temperatures the amplitudes are almost independent of the 
atomic mass, at low temperatures they increase as the mass is lowered, 
due to quantum zero-point motion. A larger amplitude {\em notices} more
effectively the anharmonicity of the interatomic potential, thereby
yielding a mass-dependence for several structural and thermodynamic
properties.

Atomistic simulations (molecular dynamics and Monte Carlo)
have been used to study structural and thermodynamic properties
of graphene.  These simulations were performed using several types
of {\em ab-initio} \cite{an12b,ch14},
tight-binding \cite{ca09c,le13,he09a},
and empirical interatomic potentials \cite{fa07,ra16,ma14,br15,lo16}.
In most cases, C atoms were treated as classical particles,
which is reliable at high temperatures, in the order or larger than
the Debye temperature of the material. For graphene, 
this temperature is rather high: $\Theta_D^{\rm out} \gtrsim$ 1000~K
for out-of-plane vibrational modes \cite{te09,po11} and
$\Theta_D^{\rm in} \gtrsim$ 2000~K for in-plane
modes \cite{te09,po12b}.

As isotopic effects on structural variables of crystalline materials
are caused by the quantum character of the atomic nuclei and
the anharmonicity of the interatomic potentials, a suitable
theoretical method to treat this problem is the Feynman path-integral
method, a powerful approach to the statistical
mechanics of many-body quantum systems at finite temperatures.
In particular, the path-integral molecular dynamics (PIMD) technique is
well established as a procedure to study many-body problems
where anharmonic effects are nonnegligible.
Path-integral simulations have been used earlier to study
isotopic effects in condensed matter, and were especially
applied to analyze the dependence of lattice parameters and
thermodynamic properties of solids upon the mass of their
constituent atoms \cite{he02,mu95,no97,he99}

In this paper, we study graphene by PIMD simulations between
50 and 1500~K, using an effective interatomic potential, the so-called 
LCBOPII, which has been found to reliably describe several structural 
and thermodynamic properties of this 2D material \cite{lo09,ra16,he16}.
This allows us to analyze in a quantitative fashion 
the influence of isotopic mass on structural properties such as
interatomic distances and area of the graphene layer.
We consider the most abundant $^{12}$C isotope, as well as
$^{13}$C and $^{14}$C.
The anharmonicity of the vibrational modes is assessed by comparing 
the atomic mean-square displacements derived from PIMD simulations 
with those obtained from a harmonic approximation.

The paper is organized as follows. In Sec.\,II, we present the general
background to study isotopic effects in structural properties of 
condensed matter. In Sec.\,III we describe the computational method 
used in our calculations. 
In Sec.\,IV  we present our results for the internal energy, and in
Sec.\,V we discuss the atomic mean-square displacements of the C atoms. 
Isotopic effects in the interatomic
distances and in the in-plane area of the graphene sheet are displayed in
Secs.~VI and VII, respectively.
The paper closed with a summary in Sec.~VIII.

\section {General background}

For a one-dimensional harmonic oscillator of frequency $\omega$ and
mass $M$, the classical mean-square displacement (MSD) at temperature 
$T$ is given by $(\Delta x)^2 = k_B T / M \omega^2$. Since the 
vibrational frequency scales with the mass as $M^{-1/2}$, then
$(\Delta x)^2$ is independent of $M$. 
In a quantum formulation, $(\Delta x)^2$ depends on $M$, and
in the ground state one has $(\Delta x)^2 = \hbar / 2 M \omega$,
so $(\Delta x)^2 \propto M^{-1/2}$, and the MSD grows for decreasing
mass. In both harmonic cases (classical and quantum), the mean position 
$\langle x \rangle$ does not change with temperature \cite{co77}. 
The same happens for a harmonic description of phonons in solids, 
so it cannot predict any thermal expansion or isotopic effects in
equilibrium structural properties \cite{ki96,as76}.

Here we are interested in the dependence of structural properties of 
graphene on isotopic mass. Such a dependence does not appear in
classical calculations, even in the presence of anharmonicity.
This is known for 3D solids, and can be 
straightforwardly seen from arguments of statistical mechanics.
The canonical partition function ($NVT$) for $N$ particles of mass
$M$ arranged in a crystalline 3D solid with volume $V$ at
temperature $T$ is given by \cite{re65,gr95}:
\begin{equation}
  Z_{NVT} = \frac{1}{h^{3N}}
     \int_{V^N} d {\bf r}^N  \int_{R^{3N}} d {\bf p}^N
     \exp \left[ - \beta  H({\bf r}^N, {\bf p}^N) \right]  \, ,
\end{equation}
where $\beta = 1 /k_B T$, ${\bf r}^N = ({\bf r}_1, ..., {\bf r}_N)$
and ${\bf p}^N = ({\bf p}_1, ..., {\bf p}_N)$ are the coordinates
and momenta of the particles, and $R$ is the real axis. 
The Hamiltonian is
\begin{equation}
  H({\bf r}^N, {\bf p}^N)  =
      \sum_{i=1}^N \frac{{\bf p}_i^2}{2 M} + U({\bf r}^N)   \, ,
\end{equation}
where the first and second term on the r.h.s. are the kinetic and 
potential energy.   In a classical model, the kinetic energy can be
integrated out to yield:
\begin{equation}
  Z_{NVT}^{\rm cl} =  \frac{1}{h^{3N}}
    \left( \frac{2 \pi M}{\beta} \right)^{3N/2}
    \int_{V^N} d {\bf r}^N  \exp \left[ - \beta  U({\bf r}^N) \right] \, .
\end{equation}
This means that equilibrium properties that depend only on the coordinates
${\bf r}^N$ are independent of the mass $M$, as this variable appears
only in the prefactor of the integral and disappears when taking average
values of functions of the coordinates. Thus, for example, the
mean-square displacement of particle $i$ is given by 
$(\Delta r_i)^2 =  \left< {\bf r}_i^2 \right> - 
  \left< {\bf r}_i \right>^2$, and does not change with the mass $M$.

In quantum statistical physics, however, positions and momenta
do not commute, so the kinetic energy part in the Hamiltonian cannot
be integrated out and the mass $M$ affects the average values of
position-dependent variables, as for example $(\Delta r_i)^2$.
The same arguments are valid for 2D materials such as graphene.
Note that in this case the integration for the space coordinates
${\bf r}^N$ is restricted to a limited area in
the graphene plane, and is unrestricted in the out-of-plane direction
(see below), but this fact does not alter our conclusion.

In summary, isotopic effects in equilibrium structural properties of 
2D and 3D materials are pure quantum effects, as they do not appear 
in classical-like calculations. Moreover, they are typical 
anharmonic effects, like thermal expansion, and do not show up 
in the absence of anharmonicity in the interatomic potentials.

As mentioned in the Introduction, several studies based on classical
atomistic simulations have been devoted to study structural, dynamic, 
and elastic properties of graphene in equilibrium conditions at finite 
temperatures \cite{fa07,lo16}. These methods are, however, useless
to study isotopic effects as they are insensitive to the atomic
mass. Then, a procedure such as those based on quantum path integrals
is necessary for this purpose.
Apart from quantum atomistic simulations, another method used earlier
to study isotopic effects in solids is the quasiharmonic approximation
(QHA) \cite{de96,ga96,he00a}. This procedure has been found to yield 
reliable results
in several studies of solids, and will be employed here to interpret 
some trends of the results of our simulations. However, a detailed 
calculation of isotopic effects using the QHA is out of the scope 
of this paper.

\section{Computational Method}

\subsection{Path-integral molecular dynamics}

In this paper we use the PIMD method to obtain equilibrium properties of 
graphene at several temperatures for different isotopic masses.
This procedure relies on the path-integral formulation of statistical
mechanics, a nonperturbative method to study many-body
quantum systems at finite temperatures \cite{fe72}.
It profits from the fact that the partition function of a quantum system 
may be expressed in a form equivalent to that of a classical one, 
resulting from replacing each quantum particle by a ring polymer made of 
$N_{\rm Tr}$ (Trotter number) classical particles ($beads$), joined 
by harmonic springs with constant 
$k_{\rm har} = M N_{\rm Tr} / \beta^2 \hbar^2$ \cite{gi88,ce95}.
Details on this simulation technique are given
elsewhere \cite{gi88,ce95,he14}.

We employ the molecular dynamics method to sample the configuration
space of the classical isomorph of our quantum system ($N$ carbon atoms).
The dynamics in this 
computational procedure is artificial, as it does not coincide with  
the quantum dynamics of the actual particles under consideration. 
This procedure is, however, well suited for adequately sampling the
many-body configuration space, yielding accurate results for 
time-independent equilibrium properties of the quantum system.

The Born-Oppenheimer surface for the nuclear dynamics is obtained from 
a long-range carbon bond-order potential, LCBOPII \cite{lo03b,lo05,gh08},
which has been used earlier to perform classical simulations of 
diamond \cite{lo05}, graphite \cite{lo05}, liquid carbon \cite{gh05},
and more recently graphene \cite{fa07,za09,za11,lo16}.
For graphene, in particular, this effective potential has been found
to give a good description of elastic properties such as 
Young's modulus \cite{za09,po12}.   Also, at 300 K it predicts 
a bending modulus $\kappa$ = 1.6 eV \cite{ra16},
close to the best fit to experimental and theoretical results
obtained by Lambin \cite{la14}.

The calculations presented here were performed in the isothermal-isobaric 
ensemble, where one fixes the number of carbon atoms ($N$), the applied 
in-plane stress (here $P = 0$), and the temperature ($T$).
We used effective algorithms for carrying out PIMD simulations
in this statistical ensemble, as those described in the
literature \cite{tu92,tu10,ma99}.
We employed staging variables to define the bead coordinates, and
the constant-temperature ensemble was generated by coupling chains
of four Nos\'e-Hoover thermostats to each staging variable.
An additional chain of four barostats was coupled to the area 
of the simulation box to yield the required zero pressure \cite{tu10,he14}.

The equations of motion have been integrated by using
the reversible reference system propagator algorithm (RESPA), 
which permits to define different time steps for the integration of 
fast and slow degrees of freedom \cite{ma96}.
For the time step associated to the interatomic forces we took
$\Delta t$ = 1 fs, which gave an adequate convergence for the
studied variables.
For the evolution of the fast dynamical variables (thermostats and 
harmonic bead interactions), we used $\delta t = \Delta t/4$, 
as in previous PIMD simulations \cite{he06,he11}.
The kinetic energy $E_{\rm kin}$ has been calculated by using 
the so-called virial estimator \cite{he82,tu10}.

The configuration space was sampled at temperatures between 
50~K and 1500~K for the carbon isotopes $^{12}$C, $^{13}$C, and
$^{14}$C. In each case, we have considered isotopically pure 
graphene. We have checked that isotopic effects corresponding
to mixtures of carbon isotopes coincide within error bars with 
the values obtained from linear interpolation of the results
for isotopically pure samples. 
To analyze the dependence of several variables on 
the isotopic mass $M$, we carried out additional PIMD simulations
for several values of $M$ which do not correspond to actual
carbon isotopes. Moreover, for comparison with the results of our 
quantum simulations, some classical molecular dynamics (MD) 
simulations were also carried out (this corresponds in our context 
to setting $N_{\rm Tr}$ = 1).
We note that the classical limit for equilibrium properties can be
also reached in the high-mass limit ($M \to \infty$), as the force 
constant $k_{\rm har}$ between beads diverges and all the beads 
corresponding to a particle (here atomic nucleus) collapse into 
a point, therefore the spatial quantum delocalization vanishes.
In our PIMD simulations the Trotter number was taken proportional 
to the inverse temperature, so that $N_{\rm Tr} \, T$ = 6000~K, 
which roughly keeps a constant precision for the PIMD results at 
different temperatures \cite{he06,he11,ra12}.

We have considered rectangular simulation cells with $N$ = 960 and 
similar side length in the $x$ and $y$ directions of the $(x, y)$ 
reference plane, and periodic boundary conditions were assumed.
For a given temperature, a simulation run consisted of
$2 \times 10^5$ PIMD steps for system equilibration, followed by
$4 \times 10^6$ steps for the calculation of ensemble average properties.
Some simulations were carried out for $N$ = 240 and 448, and the results
for the isotopic effects agreed within error bars with those found
for $N$ = 960.

\subsection{Mean-square displacements}

PIMD simulations may be used to analyze the atomic spatial delocalization
at finite temperatures. This includes thermal (classical) motion,
and a delocalization due to the quantum nature of the atomic nuclei, 
which can be measured by the spreading of the quantum paths.
For a given path, one can define the center-of-gravity
(centroid) as
\begin{equation}
   \overline{\bf r}_i = \frac{1}{N_{\rm Tr}}
          \sum_{j=1}^{N_{\rm Tr}} {\bf r}_{ij} \; ,
\label{centr}
\end{equation}
where ${\bf r}_{ij} \equiv (x_{ij}, y_{ij}, z_{ij})$ is the 3D position
of bead $j$ in the ring polymer associated to nucleus $i$.
Hence, the mean-square displacement $(\Delta r_i)^2$ 
in a PIMD run is given by
\begin{equation}
  (\Delta r_i)^2 = \frac{1}{N_{\rm Tr}} \left< \sum_{j=1}^{N_{\rm Tr}}
           ( {\bf r}_{ij} - \left< \overline{\bf r}_i \right>)^2
           \right>    \, ,
\label{deltar2}
\end{equation}
where $\langle ... \rangle$ means an ensemble average.

The quantum delocalization of a particle is related in our context
to the spread of the paths associated to it,
which can be measured by the mean-square {\em radius-of-gyration}
$Q_i^2$ of the ring polymers:
\begin{equation}
  Q_i^2 = \frac{1}{N_{\rm Tr}} \left< \sum_{j=1}^{N_{\rm Tr}}
             ({\bf r}_{ij} - \overline{\bf r}_i)^2 \right>    \, .
\label{qr2}
\end{equation}
The total MSD of nucleus $i$ at 
temperature $T > 0$ includes, apart from $Q_i^2$, another term 
corresponding to motion of the centroid $\overline{\bf r}_i$, i.e. 
\begin{equation}
    (\Delta r_i)^2 = Q_i^2 + C_i^2  \, ,
\label{deltar2b}
\end{equation}
with $C_i^2  =   \langle  \overline{\bf r}_i^2 \rangle -
\langle  \overline{\bf r}_i \rangle^2$.
This term $C_i^2$ is a semiclassical thermal contribution to 
$(\Delta r_i)^2$, as it converges at high temperature to the MSD
given by a classical model ($Q_i^2 \to 0$).

For graphene, we call $(x, y)$ the coordinates on
the plane defined by the simulation cell, and $z$ the out-of-plane
direction.  Then, we have expressions like
those given above for each direction $x$, $y$, and $z$.
For example, $Q_i^2 = Q_{i,x}^2 + Q_{i,y}^2 + Q_{i,z}^2$.
In the results presented below, we will show data for
the in-plane MSD, defined as an average for the
$N$ atoms in the simulation cell:
\begin{equation}
  (\Delta r)_p^2 = \frac1N  \sum_{i=1}^N 
	\left[ (\Delta x_i)^2 + (\Delta y_i)^2  \right]  \, .
\end{equation}
Similarly, one can define an average MSD $(\Delta z)^2$ 
in the out-of-plane direction \cite{he16}.

\section{Internal energy}

Here we present and discuss the internal energy for graphene
made up of different carbon isotopes, as derived from
our isothermal-isobaric ensemble for external stress $P = 0$
and several temperatures.
In a classical calculation, one finds for the zero-temperature limit
a planar graphene surface with an interatomic distance of 1.4199 \AA.
This corresponds to the minimum-energy configuration with the
atomic nuclei fixed on their equilibrium sites without spatial
delocalization, and defines the energy $E_0$, which is taken
as a reference for our calculations at finite temperatures.
In a quantum formulation, zero-point motion induces out-of-plane
atomic fluctuations in the low-temperature limit, and the graphene
layer is not strictly planar.
Moreover, anharmonicity of in-plane vibrations gives rise to
a zero-point bond dilation, causing an expansion of the graphene 
lattice, which will be discussed below.

\begin{figure}
\vspace{-0.6cm}
\includegraphics[width=7cm]{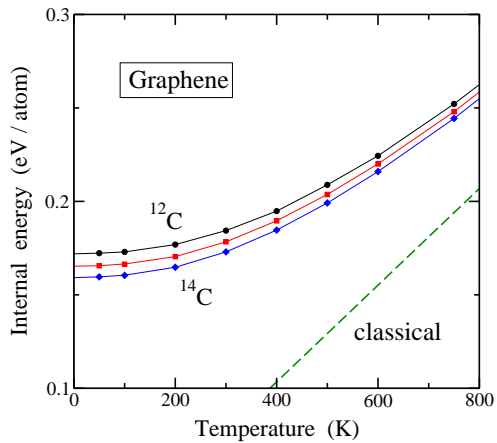}
\vspace{-0.5cm}
\caption{Temperature dependence of the internal energy of graphene
in the region up to 800 K. Symbols represent simulation results
obtained for $^{12}$C (circles), $^{13}$C (squares),
and $^{14}$C (diamonds).
Error bars are smaller than the symbol size.
Solid lines are guides to the eye.
The dashed line displays the classical vibrational energy per atom:
$E_{\rm vib} = 3 k_B T$.
}
\label{f1}
\end{figure}

In Fig.~1 we show the internal energy, $E_{\rm int}$, of graphene 
as a function of temperature for the different carbon isotopes: 
$^{12}$C (circles), $^{13}$C (squares), and $^{14}$C (diamonds), where
symbols represent results of PIMD simulations.
These simulations give separately
the potential ($E_{\rm pot}$) and kinetic ($E_{\rm kin}$) inputs
to the internal energy \cite{he82,tu10,he14},
so for $P = 0$ we have 
$E_{\rm int} = E - E_0 = E_{\rm kin} + E_{\rm pot}$.
Note that most of the internal energy corresponds to the vibrational 
energy associated to in-plane and out-of-plane modes of graphene.
A small part of the potential energy corresponds to the
{\em elastic} energy, which appears for changes in the in-plane
area of graphene, mainly at high temperatures 
(an anharmonic effect) \cite{he16}.

The internal energy for $^{12}$C is found to converge at
low $T$ to 171 meV/atom, which gives the zero-point energy of
the system. For heavier isotopes we find at low temperature
an energy shift of --7 and --13 meV/atom for $^{13}$C and $^{14}$C,
respectively.
For comparison, we also show in Fig.~1 results of the internal
energy found in classical MD simulations (dashed line).
These data are very close to the classical harmonic expectancy,
i.e., $E_{\rm int}^{\rm cl} = 3 k_B T$ per atom.
At high temperatures, the energy obtained from quantum simulations
for the different carbon isotopes converges to that of classical
simulations, but at $T$ = 800 K we still observe an appreciable
difference between quantum and classical energy values.
This is not strange, since the Debye temperature of graphene
turns out to be $\Theta_D^{\rm out} \gtrsim$ 1000~K 
for out-of-plane modes \cite{te09,po11} and
$\Theta_D^{\rm in} \gtrsim$ 2000~K for in-plane 
modes \cite{te09,po12b}, as indicated in the Introduction.

To make contact of our simulation results with a microscopic view
based on the atomic vibrations, we consider 
a harmonic approximation (HA) for the vibrational modes of graphene.
For a pure HA, the vibrational energy per atom of graphene 
is given by
\begin{equation}
 E_{\rm int}^{\rm HA}  =  \frac{1}{N} \sum_{r,\bf k}
   \frac12 \hbar \omega_r({\bf k})
   \coth \left[ \frac12 \beta \hbar \omega_r({\bf k}) \right]  \, ,
\label{evib}
\end{equation}
where the index $r$ (= 1, ..., 6) refers to the phonon bands:
two branches with atomic displacements along the $z$ direction
(ZA and ZO), and four bands with in-plane displacements
(LA, TA, LO, and TO) \cite{ra19}.
Here ${\bf k} = (k_x, k_y)$ are wavevectors in the 2D hexagonal
Brillouin zone of the reciprocal lattice associated to the
simulation cell with size $N$ \cite{mo05,ka11,ra16,ra19}.
The frequencies $\omega_r({\bf k})$ correspond to the minimum-energy
configuration of graphene (energy $E_0$), as derived from
a diagonalization of the dynamical matrix for the LCBOPII potential.
Note that for a HA of the vibrational modes, 
one has $E_{\rm kin} = E_{\rm pot}$ (virial theorem \cite{la80,fe72})
for all temperatures in both classical and quantum approaches.

In the limit $T \to 0$, one has the zero-point energy:
\begin{equation}
   E_{\rm int}^0  =  \frac{1}{N} \sum_{r,\bf k}
        \frac12 \hbar \omega_r({\bf k})  \, .
\label{evib0}
\end{equation}
Taking into account that the frequencies $\omega_r({\bf k})$
scale with isotopic mass as $M^{-1/2}$, at low $T$ the ratio 
between internal energies for masses $M_1$ and $M_2$ converges 
to $(M_2 / M_1)^{1/2}$ in the HA.  On the other side, 
in the high-temperature limit, i.e.,
$\hbar \omega_r({\bf k}) / k_B T \ll 1 $ for all 
$\omega_r({\bf k})$, we have
$E_{\rm int} \to 3 k_B T$ and the internal energy converges
to the classical value, which is independent of the isotopic mass.

\begin{figure}
\vspace{-0.6cm}
\includegraphics[width=7cm]{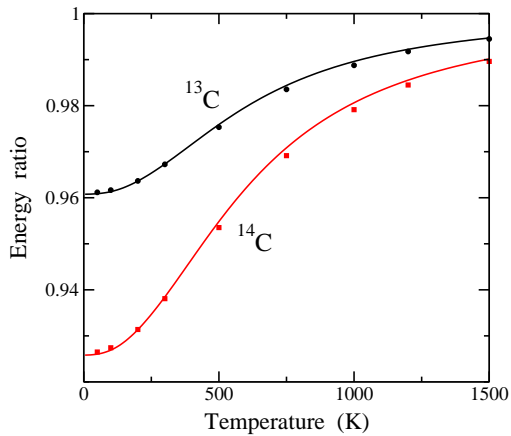}
\vspace{-0.5cm}
\caption{Ratio between the internal energy of graphene for
different carbon isotopes: $E^{13}_{\rm int} / E^{12}_{\rm int}$
(circles) and $E^{14}_{\rm int} / E^{12}_{\rm int}$ (squares).
Error bars are less than the symbol size.
Solid lines represent energy ratios corresponding to the
harmonic approximation described in text (see Eq.~(\ref{evib})).
}
\label{f2}
\end{figure}

The behavior predicted by the HA for the energy ratio between different
isotopes is basically what we find from PIMD simulations of graphene.
In Fig.~2 we present the ratio between the internal energy of graphene
for the carbon isotopes, as derived from PIMD simulations: 
$E^{13}_{\rm int} / E^{12}_{\rm int}$ (circles) and 
$E^{14}_{\rm int} / E^{12}_{\rm int}$ (squares).
For comparison, we also show as solid lines the energy ratios obtained 
by using the HA given by Eq.~(\ref{evib}).
At low temperature, the energy ratios derived from the simulations 
converge to values close to the harmonic expectancy.  
For increasing temperature, the results of the PIMD simulations 
lie below those of the HA, and both sets of results approach each
other for $^{13}$C and $^{14}$C at high $T$, as they converge to
the classical limit (energy ratio equal to unity, irrespective of
the anharmonicity).

\section{Atomic mean-square displacements}

\begin{figure}
\vspace{-0.6cm}
\includegraphics[width=7cm]{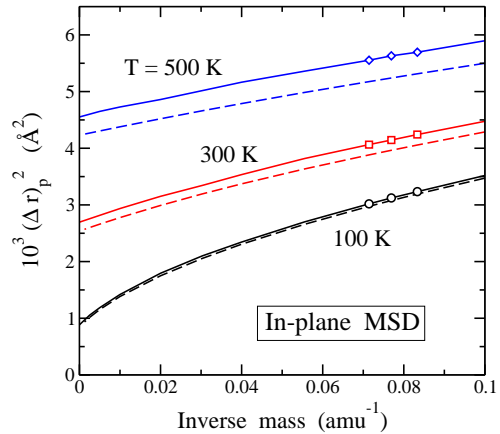}
\vspace{-0.5cm}
\caption{In-plane MSD of carbon atoms vs inverse isotopic mass
$M^{-1}$ for $T =$ 100 K (circles), 300 K (squares), and
500 K (diamonds). Solid lines were obtained from PIMD simulations
of graphene with the LCBOPII potential for $N$ = 960.
Symbols correspond to actual carbon isotopes; from left to
right: $^{14}$C, $^{13}$C, and $^{12}$C.
Dashes lines represent MSDs derived from the harmonic approximation
described in the text.    The classical limit
corresponds to $M^{-1} \to 0$.
}
\label{f3}
\end{figure}

In this section we present results for the mean-square displacements
of C atoms in graphene, yielded by PIMD simulations.
These MSDs depend on the isotopic mass, since the vibrational
amplitudes are larger for smaller mass.
In Fig.~3 we show the in-plane MSD of C atoms, $(\Delta r)_p^2$,
as a function of the inverse isotopic mass $1/M$ for 
$T$ = 100, 300, and 500~K (solid lines).
These data were obtained from PIMD simulations for different
masses $M$. Symbols correspond to the actual carbon isotopes;
from left to right: $^{14}$C, $^{13}$C, and $^{12}$C.
The limit $1/M \to 0$ (left part of the figure) corresponds
to the classical limit.

For a given isotopic mass $M$, $(\Delta r)_p^2$ increases for
rising temperature, as expected from the increase in vibrational
amplitudes. Moreover, given a temperature $T$, 
the in-plane MSD rises from the classical limit 
as the atomic mass is lowered ($1/M$ rises).
This increase is more important for lower temperature, as
shown in Fig.~3 for $T = 100$~K, due to the larger importance of
quantum effects at lower $T$.

The in-plane MSD at temperature $T$ can be written in a HA as
\begin{equation}
 (\Delta r)_p^2 \approx \frac1N \sum_{r,{\bf k}} 
   \frac{\hbar}{2 M \omega_r({\bf k})}
   \coth \left( \frac{\hbar \omega_r({\bf k})}{2 k_B T} \right)  \, .
\label{drp2}
\end{equation}
Here the index $r$ (= 1, ..., 4) runs over the phonon bands
with in-plane atomic displacements (LA, TA, LO, and TO) \cite{ra19}.
Results of the HA are displayed in Fig.~3 as dashed lines.
At $T = 100$~K these data follow closely those derived from PIMD 
simulations, and both sets of results depart one from the other 
at higher temperatures. This indicates an increase in the effect of
anharmonicity on the in-plane MSD. The results of the simulations
become larger than the prediction of the pure HA, but this
approximation yields reasonable results at temperatures in the order 
of 100~K.   The classical limit for the MSD is given in a HA by:
\begin{equation}
    C_p^2 \approx \frac1N \sum_{r,{\bf k}} \frac{k_B T}
	   {M \omega_r({\bf k})^2}  \, ,
\label{cp2}
\end{equation}
and corresponds to the limit of $(\Delta r)_p^2$ on the left part
of Fig.~3. In this classical approximation, $C_p^2$ is
proportional to the temperature $T$.

For the out-of-plane displacements ($z$ direction), it has been shown 
earlier that the character of the motion is predominantly 
classical above a crossover temperature $T_c$, which depends on 
the system size and decreases as $N$ is raised \cite{he16}.
This occurs because $Q_z^2$ decreases for rising temperature, 
while $C_z^2$ increases almost linearly with $T$, and the slope
of $C_z^2(T)$ grows with system size $N$ \cite{he16}. 
Increasing $N$ gives rise to the onset of vibrational 
modes with longer wavelength $\lambda$. In fact, one can define an 
effective cut-off $\lambda_{max} \approx L$, with $L = (N A_p)^{1/2}$
($A_p$: in-plane area per atom, see below).
Calling $k = |{\bf k}|$, this translates into
$k_{min} = 2 \pi / \lambda_{max}$, i.e., $k_{min} \sim N^{-1/2}$.

For $N = 960$, the crossover for the out-of-plane motion occurs 
at $T_c \approx 10$~K \cite{he16},
so for the temperatures considered here the atomic MSD in the
$z$ direction is basically controlled by the 
classical contribution $C_z^2$, i.e., we have $C_z^2 \gg Q_z^2$.
This is mainly due to low-frequency out-of-plane modes
({\em flexural} ZA band), which cause large vibrational amplitudes, 
and behave classically at the temperatures considered here.
Thus, at a given temperature, changes in $(\Delta z)^2$ for 
the different carbon isotopes
are expected to be negligible, since classical-like motion
yields MSDs independent of the mass.
We have checked this from the results of our PIMD simulations,
and found that differences in the out-of-plane MSDs for different
carbon isotopes are less than the statistical error bars.
This means that, at least for the temperatures considered here 
($T \geq 50$~K), the isotopic effects in the interatomic distances
and in-plane area presented below are basically controlled by
the atomic in-plane motion.

\section{Interatomic distances}

\begin{figure}
\vspace{-0.6cm}
\includegraphics[width=7cm]{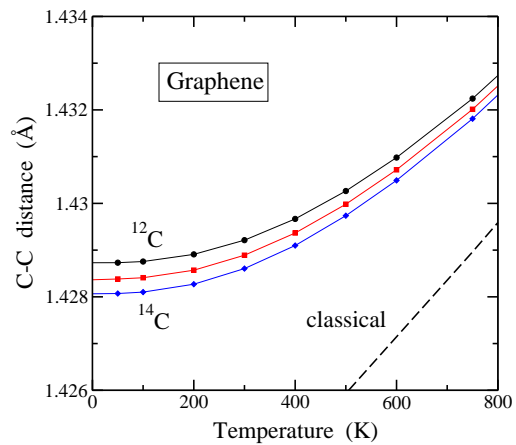}
\vspace{-0.5cm}
\caption{Temperature dependence of the interatomic C--C distance
in graphene for $^{12}$C (circles), $^{13}$C (squares),
and $^{14}$C (diamonds). Solid lines are guides to the eye.
The dashed line represents results for $d_{\rm C-C}$ derived from
classical MD simulations.
}
\label{f4}
\end{figure}

In this section we study changes of interatomic distances in graphene
with temperature and isotopic mass.
The temperature dependence of the equilibrium C--C distance,
$d_{\rm C-C}$, derived from PIMD simulations
is shown in Fig.~4 for the different carbon isotopes: $^{12}$C (circles), 
$^{13}$C (squares), and $^{14}$C (diamonds).
For $^{12}$C we find in the low-temperature limit an interatomic
distance of 1.4287 \AA, which grows as temperature is raised, as
expected for the thermal expansion of the graphene sheet.
The interatomic distance corresponding to the minimum-energy
configuration (flat graphene sheet with energy $E_0$) amounts 
to 1.4199 \AA, which means that the zero-point expansion of the C--C 
bond for $^{12}$C is $(\Delta d)_0 = 8.8 \times 10^{-3}$~\AA, 
i.e., a 0.6\% of the classical result. 
This value is larger than the whole vertical axis presented in Fig.~4.

The dashed line in Fig.~4 corresponds to the results of classical 
simulations, which display a dependence nearly linear with temperature.
A comparison between results derived from classical MD and PIMD
simulations of graphene was presented in Ref.~\cite{he16}.
The bond expansion due to quantum zero-point motion was found to be
in the order of the thermal expansion predicted by the classical model
from $T$~=~0 to 800~K.   In the PIMD results,
the rise in interatomic distance from $T$ = 0 to 300~K is small,
amounting to $5 \times 10^{-4}$ \AA,
about 15 times smaller than the zero-point expansion $(\Delta d)_0$.
The size effect of the finite simulation cell is negligible
for our purposes \cite{he16}. For some selected temperatures, we have
checked that results for $d_{\rm C-C}$ obtained
for $N$ = 1560 coincide within error bars with those presented
here for $N = 960$.

The results for interatomic distances derived from classical and 
quantum simulations for a single layer of graphene are qualitatively 
similar to those found earlier for other carbon-based materials.  
For example, the zero-point expansion of the C--C bond 
for diamond was found to be $(\Delta d)_0 = 7.4 \times 10^{-3}$~\AA, 
i.e., a 0.5\% of the classical prediction \cite{he00c}.

\begin{figure}
\vspace{-0.6cm}
\includegraphics[width=7cm]{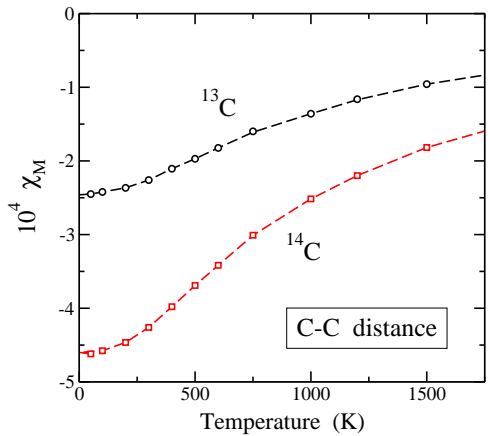}
\vspace{-0.5cm}
\caption{Temperature dependence of $\chi_M = (d_M - d_{12}) / d_{12}$
for $^{13}$C (circles) and $^{14}$C (squares), as derived from
PIMD simulations of graphene.
Dashed lines are guides to the eye.
Error bars are in the order of the symbol size.
}
\label{f5}
\end{figure}

For graphene, the C--C distance derived from PIMD simulations is 
found to decrease for increasing isotopic mass (see Fig.~4).
At $T$ = 50 K it is reduced by 3.5 and $6.6 \times 10^{-4}$~\AA\ for 
$^{13}$C and $^{14}$C, respectively, with respect to the C--C distance
in $^{12}$C. These differences are reduced for rising temperature,
as nuclear quantum effects become less relevant.
In Fig.~5 we display the temperature dependence of the ratio 
$\chi_M = (d_M - d_{12}) / d_{12}$, where $d_M$ is 
the interatomic distance corresponding to isotopic mass $M$.
For $^{13}$C and $^{14}$C, this ratio converges for $T \to 0$ to 
$-2.5 \times 10^{-4}$ and $-4.6 \times 10^{-4}$, respectively.

For diamond, it was found at low $T$ a ratio 
$\chi_{13} = -1.8 \times 10^{-4}$ for $^{13}$C, somewhat 
smaller than our result for graphene.  
This is most probably due to the different
anharmonicity associated to the hybridization of carbon atoms 
in both materials: sp$^3$ in diamond vs sp$^2$ in graphene.
In fact, the thermal expansion of the interatomic distance
(or lattice parameter) for diamond from $T$ = 0 to 1000~K amounts
to a 0.2\%, vs a 0.4\% for the expansion of $d_{\rm C-C}$ in
graphene \cite{he00c}.

The isotopic effect in the C--C distance studied here is
clearly larger than the present sensitivity of diffraction
techniques. Thus, for example, isotopic effects in lattice 
parameters and interatomic distances of crystalline materials 
can be measured with high precision using x-ray standing 
waves \cite{ka98}.

The zero-point dilation $(\Delta d)_0$ and the thermal 
bond expansion at $T > 0$ presented here are a consequence 
of anharmonicity in the interatomic potential, 
similar to 3D crystalline solids, e.g., diamond (see above). 
For graphene, this is mainly caused by anharmonicity in the
stretching vibrations of the sp$^2$ C--C bond. A more complex
anharmonic effect is the thermal variation of the graphene 
in-plane area, for the coupling between in-plane and
out-of-plane vibrational modes, as discussed in Sec.~VII.

In this section, we have analyzed the interatomic distances
in isotopically pure graphene. As mentioned in Sec.~III.A, one
expects that structural variables for mixtures of carbon isotopes
can be obtained from linear interpolation between those corresponding
to isotopically pure samples. To check this point we have
carried out some PIMD simulations for carbon mean mass of
12.25, 12.5, and 12.75 amu, and found that the results for the
C--C distance coincide within error bars with those given
from interpolation between the data obtained for $M$ = 12 and
13 amu. Something similar happens for the in-plane area $A_p$
discussed in the following section.
In connection with this,
we note that to calculate several properties of 3D crystals
with isotopically mixed composition, it has been usually
assumed that each atomic nucleus in the material has a mass
equal to the average mass.
This was called {\em virtual-crystal approximation}
\cite{de96,he99,ca05b,he09c}, and it has been found to
give results that coincide with those yielded by simulations
in which real isotopic mixtures are considered \cite{he09c}.

\section{Layer area}

The simulations presented here have been performed
in the isothermal-isobaric ensemble, as indicated in Sec.~III.A.
Thus, for a given isotopic mass $M$, in a simulation run we fix 
the number of carbon atoms $N$, the temperature $T$, 
and the applied stress in the $(x, y)$ plane (here $P = 0$), 
allowing for variations in the area of the simulation cell.
In the following we will call $A_p$ the in-plane area per C atom.

In the simulations, carbon atoms are allowed to freely move in 
the $z$ coordinate (out-of-plane direction), so that the {\em real} 
surface of a graphene layer is not strictly planar, and has an area 
(in 3D space) larger than the in-plane area \cite{ha16,he16}.
The difference between the {\em real} area $A$ and the {\em in-plane}
area $A_p$ (called $excess$ \cite{he84,fo08,he20} or 
$hidden$ area \cite{ni17}) has been discussed 
earlier for biological membranes \cite{im06,wa09,ch15}, and 
more recently for crystalline
membranes such as graphene \cite{ra17,he20}.
The {\em real} area $A$ can be calculated from PIMD simulations of 
graphene by a triangulation based on the atomic positions 
\cite{ra17,he19}.
This area $A$ is larger than $A_p$, and the difference between both
increases with temperature, since the actual surface becomes
increasingly bent as temperature is raised and out-of-plane atomic 
displacements around the reference plane are larger.
Here we will concentrate on the in-plane area $A_p$, as it is
the variable conjugate to the external pressure $P$.

\begin{figure}
\vspace{-0.6cm}
\includegraphics[width=7cm]{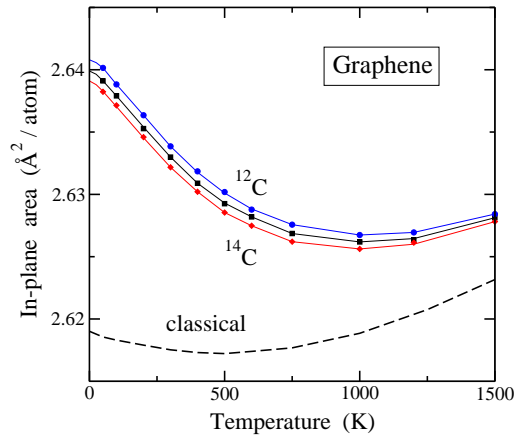}
\vspace{-0.5cm}
\caption{In-plane area as a function of temperature for graphene made
of different carbon isotopes: $^{12}$C (circles), $^{13}$C (squares),
and $^{14}$C (diamonds). These results were obtained from PIMD
simulations for $N$ = 960.
The dashed line indicates the result for $A_p$ obtained in
classical MD simulations.
}
\label{f6}
\end{figure}

In Fig.~6 we show the temperature dependence of $A_p$, 
as derived from PIMD simulations for three carbon isotopes:
$^{12}$C (circles), $^{13}$C (squares), and $^{14}$C (diamonds).
In the three cases, $A_p$ decreases for rising $T$, reaches a minimum
at a temperature $T_m \sim 1000$~K, and increases at higher temperature.
This behavior was found and discussed earlier for $^{12}$C \cite{he16}.
For comparison with the results of PIMD simulations, we also show 
in Fig.~6 the temperature dependence of $A_p$ derived from classical
MD simulations (dashed line).
In the limit $T \to 0$, the difference between classical and 
quantum data for $^{12}$C converges to 0.022 \AA$^2$/atom. 
This difference decreases for rising temperature,
as nuclear quantum effects become less important.
In particular, it is relevant that the decrease in $A_p$ obtained from
PIMD simulations from T = 0 to 1000~K is much larger than that 
found from classical molecular dynamics
and Monte Carlo simulations \cite{he16,za09,ga14,br15}.

Given a temperature, the in-plane area $A_p$ decreases for rising 
isotopic mass $M$, and eventually reaches the classical limit 
for large $M$.
For $T = 50$~K, the decrease in $A_p$ with respect to $^{12}$C graphene
amounts to $\Delta A_p = -1.0$ and $-1.9 \times 10^{-3}$ \AA$^2$/atom,
for $^{13}$C and $^{14}$C, respectively. For the classical limit
($M \to \infty$), the shift is $-2.1 \times 10^{-2}$ \AA$^2$/atom.

The behavior of $A_p$ as a function of temperature may be explained
as a result of two competing factors.
First, the real area $A$ increases as temperature is raised, as 
happens for the interatomic distance $d_{\rm C-C}$ shown in Fig.~4. 
Second, bending (rippling) of the graphene surface
gives rise to a reduction of its 2D projection, i.e., $A_p$.
At low temperature, this decrease due to out-of-plane motion dominates
over the thermal expansion of the actual surface (which is small at
low $T$), and then one has $d A_p / d T < 0$.
At high temperature, the rise of the real area $A$ dominates over 
the decrease in the projected area due to bending, and therefore
$d A_p / d T > 0$.
These arguments are similar to those given earlier to explain the
results of classical simulations of graphene \cite{ga14,mi15b}.

\begin{figure}
\vspace{-0.6cm}
\includegraphics[width=7cm]{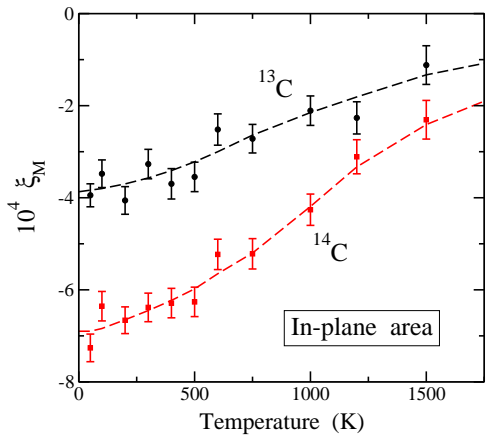}
\vspace{-0.5cm}
\caption{Temperature dependence of
$\xi_M = (A_p^M - A_p^{12}) / A_p^{12}$
for $^{13}$C (circles) and $^{14}$C (squares), as
derived from PIMD simulations of graphene.
Dashed lines are guides to the eye.
}
\label{f7}
\end{figure}

Changes in interatomic distances and in the graphene area 
(both $A$ and $A_p$) are important anharmonic effects. 
On one side, changes in the distance
$d_{\rm C-C}$ and in the real area $A$ are mainly due to anharmonicity
of the C-C stretching vibration. 
Moreover, out-of-plane vibrations cause fluctuations of the in-plane
area and a decrease in its mean value.

In Fig.~7 we present the temperature dependence of the ratio
$\xi_M = (A_p^M - A_p^{12}) / A_p^{12}$, where $A_p^M$ is 
the in-plane area for isotopic mass $M$.
For $^{13}$C and $^{14}$C, this ratio converges at low $T$ to
$3.9(2) \times 10^{-4}$ and $6.9(2) \times 10^{-4}$, respectively.
We note that the error bars in this case are clearly larger than
for the ratio $\chi_M$ shown in Fig.~5  (not displayed for 
$\chi_M$, as they are in the order of the symbol size). 
This is due to the larger statistical noise in $A_p$ than 
in $d_{\rm C-C}$ (or in the real area $A$), because of the larger
fluctuations of the former at a given temperature.
These fluctuations of $A_p$ are due to the sluggish out-of-plane 
bending modes with the largest wavelengths (smallest $k$)
and lowest vibrational frequencies corresponding to the considered
simulation cell.

It is interesting to analyze the dependence of $A_p$ on isotopic mass,
from the actual carbon isotopes to the classical limit ($M \to \infty$), 
which is in fact obtained by carrying out classical MD simulations.
This crossover can be actually visualized by plotting the in-plane area
vs the inverse isotopic mass, $1 / M$, at different temperatures,
as displayed in Fig.~8. From top to bottom, lines represent $A_p$ 
at several temperatures: 100, 300, 500, and 1000~K.
In this plot, the left side $1/M \to 0$ corresponds to the classical 
limit, and symbols indicate the actual carbon isotopes:
from left to right, $^{14}$C, $^{13}$C, and $^{12}$C.
Note that the line corresponding to $T = 1000$~K crosses the other
lines for large $M$, since in the classical limit the minimum of
$A_p$ vs $T$ appears at about 500~K \cite{he16}.
Note that the difference in $A_p$ between $^{12}$C graphene and
the classical limit changes from $2.1 \times 10^{-2}$ \AA$^2$ 
at 100~K to $7.8 \times 10^{-3}$ \AA$^2$ at 1000~K, in agreement 
with the results shown in Fig.~6.

\begin{figure}
\vspace{-0.6cm}
\includegraphics[width=7cm]{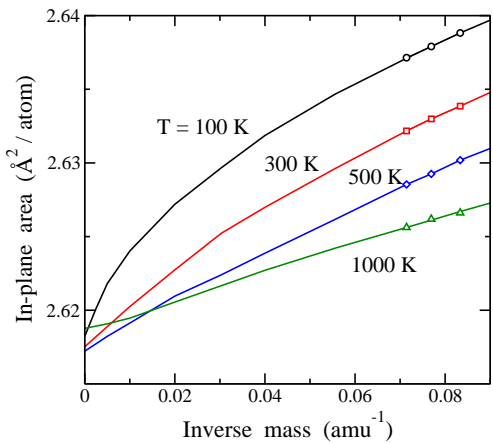}
\vspace{-0.5cm}
\caption{In-plane area of graphene vs the inverse atomic mass
$M^{-1}$ for different temperatures: 100 K (circles), 300 K (squares),
500 K (diamonds), and 1000 K (triangles). The classical limit
corresponds to $M^{-1} \to 0$.
}
\label{f8}
\end{figure}

The results of our PIMD simulations can be rationalized by
analyzing the trends expected from a 
QHA for the vibrational modes in graphene.
This method has been used earlier to study isotopic effects in
structural properties of 3D solids \cite{de96,ga96},
and more recently in graphene \cite{mo05}.
In this approach, the frequencies $\omega_r({\bf k})$ are assumed 
to change with the in-plane area, and for a given $A_p$ the modes 
are treated as harmonic vibrations.
Following these assumptions, and considering a reference mass 
$M_{\rm ref}$ (which is taken here as $M_{\rm ref}$ = 12 amu), 
the difference between in-plane areas for isotopic masses 
$M$ and $M_{\rm ref}$ at $T = 0$,
$\Delta A_p(0) = A_p^M(0) - A_p^{\rm ref}(0)$,
is given by the expression (see the Appendix):
\begin{equation}
  \Delta A_p(0)  = - \frac12  \left[  A_p^{\rm ref}(0) - A_0  \right]
      \, \frac {\Delta M}{M_{\rm ref}}   \, ,
\label{dap1}
\end{equation}
with $\Delta M = M - M_{\rm ref}$.
This means that the low-temperature changes of $A_p$ due to
isotopic mass may be quantitatively explained from the change
in $A_p$ caused by zero-point motion for the reference mass,
i.e., from the difference $A_p^{\rm ref}(0) - A_0$.
Taking for this difference the value 0.022 \AA$^2$ (see above),
Eq.\,(\ref{dap1}) yields  $\Delta A_p = -9.2 \times 10^{-4}$~\AA$^2$ 
and $-1.8 \times 10^{-3}$~\AA$^2$ for $^{13}$C and $^{14}$C, respectively.
These values are very close to the low-temperature results found from
direct calculation of the in-plane area corresponding to
different atomic masses: $-1.0$ and $-1.8 \times 10^{-3}$~\AA$^2$,
respectively.

One can also find an expression for the isotopic effect on $A_p$ 
at high temperature ($T \gtrsim \Theta_D$, see the Appendix):
\begin{equation}
        \Delta A_p(T) =
        - \left[ A_p^{\rm ref}(T) - A_p^{\rm cl}(T) \right]
          \, \frac {\Delta M}{M_{\rm ref}}   \, .
 \label{dap2}
\end{equation}
This formula corresponding to high temperature is analogous 
to that given above for $T = 0$ [see Eq.~(\ref{dap1})],
the only difference between them being a factor 2 which appears
in the denominator of the low-temperature formula.
Expressions similar to Eq.~(\ref{dap2}) have been used to
analyze the isotopic effect in the volume and interatomic distances 
in 3D solids such as silicon \cite{he00a}, at temperatures higher
than the Debye temperature of the material. For graphene, we have
for in-plane vibrational modes $\Theta_D^{\rm in} \gtrsim 2000$~K, 
a temperature higher than those studied here.
In any case, at $T = 1500$~K, Eq.~(\ref{dap2}) gives 
$\Delta A_p = -4.4$ and $-8.8 \times 10^{-3}$ \AA$^2$ for
$^{13}$C and $^{14}$C, respectively, which turn out to be
larger than those found directly from the PIMD simulations,
i.e., $-3.0$ and $-6.1 \times 10^{-3}$ \AA$^2$, which confirms
that the high-$T$ approximation Eq.~(\ref{dap2}) overestimates 
the isotopic effect at $T = 1500$~K $< \Theta_D^{\rm in}$. 

We finally note that the QHA yields reasonable predictions for 
the isotopic effect on the in-plane area, specially at low 
temperatures.   Nevertheless, anharmonic force constants not included 
in this approximation can give rise to some anomalies in its 
predictions \cite{mo05,mi15b}.

\section{Summary}

PIMD simulations are a powerful tool to study isotopic effects 
in condensed matter. 
This method allows one to study phonon-related properties, further 
than harmonic approximations for the vibrational modes, and
explore their dependence on the mass of the constituent atoms,
which appears as an input parameter in the calculations. 

Here we have demonstrated the applicability of the PIMD method 
to analyze the dependence of structural properties of 2D materials
on the atomic mass. In particular, we have studied the change of
interatomic distances and in-plane area for different carbon
isotopes in a wide range of temperatures, from $T$ = 50 to 1500 K.

At low temperature, we find a relative change of the C--C distance 
in $^{13}$C and $^{14}$C graphene of $-2.5$ and $-4.6 \times 10^{-4}$ 
with respect to $^{12}$C, respectively.
For the in-plane area $A_p$, the corresponding relative changes 
amount to $-3.9$ and $-6.9 \times 10^{-4}$.
The magnitude of these isotopic effects decreases as temperature
rises. Thus, for $A_p$ in $^{13}$C graphene we find a relative
variation of $-3.4$ and $-2.1 \times 10^{-4}$ at 300 and 1000 K,
respectively,

An interesting application of path-integral simulations is to
study the evolution of structural properties of 2D materials
by changing the atomic mass from the actual isotopic masses to
the classical limit ($M \to \infty$), which is usually considered
in atomistic simulations of this type of materials.
This has been presented here for the in-plane MSD and in-plane 
area $A_p$ of graphene at several temperatures in Figs.~3 and 8.
A similar mass dependence is obtained for the interatomic 
C--C distance.

The classical limit is indeed useful to study mechanical and
vibrational properties of graphene, but nuclear quantum effects
have to be taken into account for a proper quantification of
structural and thermodynamic variables. Classical simulations 
are, in fact, insensitive to the atomic mass and thereby
to the isotopic effects discussed here. 

Apart from isotopic effects in structural properties,
the extent of anharmonicity has been quantified by comparing 
the internal energy and in-plane vibrational amplitudes with those 
obtained from a harmonic approximation.
This is particularly observable in the in-plane MSD, 
$(\Delta r)_p^2$, shown in Fig.~3 with an increasing departure 
from the harmonic calculation for rising temperature.
For $^{12}$C, the difference between PIMD and HA results is 
about 1\% at $T$ = 100~K and amounts to a 7\% at 500~K.

Path-integral simulations similar to those presented here can be useful
to analyze isotopic effects in structural and vibrational properties 
of graphane (i.e., hydrogenated graphene), where the light mass of hydrogen
can cause a larger departure of harmonicity than that found here
for graphene. 
Another interesting topic may be the dependence of the isotopic
effects discussed here on tensile external stress, which could be
also studied in the isothermal-isobaric ensemble.

\begin{acknowledgments}
This work was supported by Ministerio de Ciencia e Innovaci\'on
(Spain) through Grants FIS2015-64222-C2 and PGC2018-096955-B-C44.
\end{acknowledgments}  

\vspace{5mm}

{\bf Author contribution statement}   \\

All authors contributed equally to the paper


\appendix

\section{Quasiharmonic approximation}

In a QHA for 3D solids, the vibrational modes are described 
by harmonic oscillators of frequencies $\omega_r({\bf k})$,
which depend on the volume of the crystal \cite{de96,ga96,he00a,mo05}. 
For graphene, the in-plane area $A_p$ plays the role of the volume 
in 3D crystals. Then, the equilibrium $A_p$ for
isotopic mass $M$ at temperature $T$ and zero external stress 
can be derived by a minimization of the Helmholtz free energy \cite{mo05}.
One finds
\begin{equation}
A_p^M(T) = A_0 + \frac{1}{N B_0}
   \sum_{r, {\bf k}}  \gamma_r({\bf k})  E_r({\bf k},T)
    \hspace{0.2cm}  ,
\label{apt}
\end{equation}
where
\begin{equation}
 E_r({\bf k},T) =  \frac{1}{2} \hbar \omega_r({\bf k}) 
  \coth \left( \frac{\hbar \omega_r({\bf k}) } {2 k_B T} \right)  \, .
\label{ejt}
\end{equation} 
Here, $\omega_r({\bf k})$ are the frequencies of the phonon band $r$, 
$B_0$ is the 2D modulus of hydrostatic compression \cite{be96b}, 
$A_0$ is the zero-temperature in-plane area
in the limit of infinite isotopic mass (classical limit,
$A_0 = A_p^{\rm cl}(0)$), and
\begin{equation}
  \gamma_r({\bf k}) = - \left. \frac {\partial \ln \omega_r({\bf k})}
     {\partial \ln A_p} \right|_0
\end{equation}
is the Gr\"uneisen parameter of mode $r, {\bf k}$ \cite{as76}. 

At $T = 0$ the difference $A_p^{M}(0) - A_0$ is given by
\begin{equation}
  A_p^{M}(0) - A_0   = \frac{1}{2 N B_0}
     \sum_{r, {\bf k}}  \hbar \omega_r({\bf k}) \gamma_r({\bf k})  \, .
\label{ap1}
\end{equation}
A first-order expansion for the in-plane area as
a function of isotopic mass yields:
\begin{equation}
   A_p^{M}(0) = A_p^{\rm ref}(0) + 
     \left. \frac{\partial A_p^M(0)}{\partial M}
           \right|_{M_{\rm ref}} \Delta M   \,  ,
\label{ap2}
\end{equation}
with $\Delta M = M - M_{\rm ref}$ the mass increment with respect
to a reference mass, which is taken here as $M_{\rm ref}$ = 12 amu. 
Taking into account that the frequencies $\omega_r({\bf k})$ 
scale with isotopic mass as $M^{-1/2}$, we have
\begin{equation}
   \frac {\partial \omega_r({\bf k})} {\partial M} =
       - \frac {\omega_r({\bf k})} {2 M}  \,  ,
\label{pmk}
\end{equation}
and
\begin{equation}
   \frac {\partial A_p^{M}(0)} {\partial M} =
      - \frac {1} {2 M}  \left[  A_p^{M}(0) - A_0  \right] \, .
\label{ap3}
\end{equation}
Thus, at $T = 0$ one has for the change of in-plane area 
due to isotopic mass:
\begin{equation}
  \Delta A_p(0)  = - \frac12  \left[  A_p^{\rm ref}(0) - A_0  \right]
      \, \frac {\Delta M}{M_{\rm ref}}   \, .
\label{dap1b}
\end{equation}

At high temperatures, the difference
$A_p^M(T) - A_p^{\rm ref}(T)$ can be also expressed as a function of
the difference $A_p^{\rm ref}(T) - A_p^{\rm cl}(T)$ between the in-plane
area for the reference mass $M_{\rm ref}$ and that corresponding to the
classical limit ($M \to \infty$) at temperature $T$.
In a QHA, we have:
\begin{equation}
  A_p^M(T) - A_p^{\rm cl}(T) =  \frac{1}{N B_0}
         \sum_{r, {\bf k}} \gamma_r({\bf k})
         \left[  E_r({\bf k},T) - k_B T \right]   \, .
\label{ap4}
\end{equation}
Using a high-temperature expansion for the energy $E_r({\bf k},T)$
given in Eq.\,(\ref{ejt}), one finds for $T \gtrsim \Theta_D$:
\begin{equation}
  A_p^M(T) - A_p^{\rm cl}(T)  =   \frac{1}{12 N B_0 k_B T}
        \sum_{r, {\bf k}} \gamma_r({\bf k})
        \left[ \hbar \omega_r({\bf k}) \right]^2   \,  ,
\label{ap5}
\end{equation}
which indicates that the isotopic effect on the in-plane area
decreases for rising temperature as $1/T$.

From  Eq.\,(\ref{ap5}), and using a first-order expansion
for $\Delta A_p(T)$ as a function of $\Delta M$, analogous to that 
given above for $T = 0$ in Eq.\,(\ref{ap2}), one finds
\begin{equation}
	\Delta A_p(T) = 
	- \left[ A_p^{\rm ref}(T) - A_p^{\rm cl}(T) \right]
          \, \frac {\Delta M}{M_{\rm ref}}   \, .
 \label{dap2b}
\end{equation}
This expression corresponding to high temperature
is similar to that found for $T = 0$ [see Eq.\,(\ref{dap1b})].
The difference between them is a factor 2 appearing
in the denominator of the low-temperature expression.



\end{document}